\documentclass[pra,superscriptaddress,twocolumn]{revtex4}
\usepackage{amssymb}
\usepackage{amsmath}
\usepackage{amsfonts}
\usepackage{fixmath}
\usepackage{graphicx}
\usepackage{mathrsfs}

\newcommand{\ket}[1]{|#1\rangle}
\newcommand{\bra}[1]{\langle#1|}
\newcommand{\kket}[1]{|#1\rangle\!\rangle}
\newcommand{\bbra}[1]{\langle\!\langle#1|}
\newcommand{\bbrakket}[2]{\langle\!\langle#1|#2\rangle\!\rangle}
\newcommand{\Tr}{\textrm{Tr}}

\renewcommand{\vec}[1]{\boldsymbol{#1}}
\renewcommand{\t}[1]{\textrm{#1}}

\def\map#1{{\mathscr{#1}}}
\def\set#1{{\sf #1}}
\def\sH{\set{H}}\def\sS{\set{S}}
\def\gG{\vec{G}}
\def\Bndd#1{\set{Lin}(#1)}
\def\Choi#1{\operatorname{Choi}({\map{#1})}}
\begin{document}
\title{Quantum state decorrelation}

\author{G. M. D'Ariano}

\affiliation{QUIT Group,
  University of Pavia and CNISM, via Bassi 6, I-27100 Pavia,
  Italy.}

\author{R. Demkowicz-Dobrza\'nski}

\affiliation{Institute of Physics, Nicolaus Copernicus University, ul. Grudziacka 5, 87-100 Toru\'{n}, Poland.}

\author{P. Perinotti}

\affiliation{QUIT Group,
  University of Pavia and CNISM, via Bassi 6, I-27100 Pavia,
  Italy.}

\author{M. F. Sacchi}

\affiliation{QUIT Group,
  University of Pavia and CNISM, via Bassi 6, I-27100 Pavia,
  Italy.}

\affiliation{CNR - Istituto Nazionale per la Fisica della Materia,
Unit\`a di Pavia, Italy.}

\begin{abstract}
  We address the general problem of removing correlations from quantum
  states while preserving local quantum information as much as
  possible. We provide a complete solution in the case of two qubits,
  by evaluating the minimum amount of noise that is necessary to
  decorrelate covariant sets of bipartite states. We show that two
  harmonic oscillators in arbitrary Gaussian state can be decorrelated
  by a Gaussian covariant map.  Finally, for finite-dimensional
  Hilbert spaces, we prove that states obtained from most cloning
  channels (e.g., universal and phase-covariant cloning) can be
  decorrelated only at the expense of a complete erasure of
  information about the copied state. More generally, in finite
  dimension, cloning without correlations is impossible for continuous
  sets of states.  On the contrary, for continuos variables cloning, a
  slight modification of the customary set-up for cloning coherent
  states allows one to obtain clones without correlations.
\end{abstract}
\date\today
\maketitle

\section{Introduction}
The processing of \emph{quantum information} is subjected to a number of restrictions imposed by the
laws of quantum mechanics, which forbid basic tasks as state cloning \cite{Wootters82}, or the
universal-NOT gate \cite{Buzek99}.  Such limitations, however, are sometimes proved useful for
applications, e.~g. the no-cloning theorem, which is at the core of quantum cryptography, since it
prevents an eavesdropper from creating perfect copies of a transmitted quantum state.  Moreover, the
study of these no-go theorems allows us to broaden our understanding of quantum mechanics itself.

In a recent Letter \cite{our1} we have posed the following question: ``Is there any intrinsic
limitation in removing correlations between quantum systems?'' We are interested in the possibility
of \emph{decorrelating quantum states} nontrivially, while keeping some local information encoded on
each system. Notice that, although extensive studies has been carried out on the separability
problem, in order to distinguish classical correlation from entanglement, very little was known
before on the problem of decorrelability of quantum states. Linearity of quantum mechanics forbids
exact decorrelation of a unknown density matrix \cite{Terno}, i.~e. there exists no quantum channel
that can map an unknown multipartite quantum state to the tensor product of its local reduced
density matrices.  What about then unfaithful decorrelation that allows some additional noise on the
output decorrelated local states, and what about if the input state is not completely unknown, i.~e.
it is drawn from a smaller set of states, such as a set with some symmetry? Other questions that are
naturally raised are: how decorrelable are the states from optimal universal cloning? Is it possible
to approximately clone without correlating the copies? Is the infinite dimensional case (continuous
variables) analogous to the finite dimensional one (qudits)? In the mentioned Letter Ref.
\cite{our1} we answered to these questions.

The following facts about cloning and state estimation motivate further the interest in the problem
of quantum state decorrelation. We know that quantum information cannot be copied or broadcast
exactly, due to the no-cloning theorem.  Nevertheless, one can find approximate optimal cloning
channels which increase the number of copies of a state at the expense of the quality.  In the
presence of noise, however, (i. e. when transmitting ``mixed'' states), it can happen that we are
able to increase the number of copies without loosing the quality, if we start with sufficiently
many identical originals.  Indeed, it is even possible to {\em purify} in such a broadcasting
process---the so-called {\em super-broadcasting} \cite{our,broad}.  Clearly, a larger number of
copies cannot increase the available information about the original input state, and this is due to
the fact that the final copies are not statistically independent, and the correlations between them
limit the extractable information \cite{estcor}.  It is now natural to ask if we can remove such
correlations and make the output systems independent again.
Clearly, such quantum decorrelation cannot be achieved exactly, otherwise we would increase the
information on the state. A priori it is not excluded, however, that it is possible to decorrelate
clones at the expense of introducing some additional noise---such that state estimation fidelity
after decorrelation is not greater than before. One of the results of this paper is that clones
obtained by most cloning machines (e.g. universal, covariant) cannot be decorrelated even within
this relaxed condition (see Sec. \ref{sec:cloning}). This also implies that the non-increasing of
distinguishability of states \emph{is not} in general a sufficient condition for decorrelability.
Apart from this negative result, we will provide examples of sets of states for which decorrelation
is possible, and calculate the optimal local noise that needs to be added to achieve the task.

After review and further discussing the results of Ref. \cite{our1} with a thorough derivation, we
present new general results on the state-decorrelation problem. We will prove that for qudits
uncorrelated cloning is impossible, even probabilistically, for any set of states containing a
finite arch of states of the form $|\phi \rangle =\sqrt{p} |0\rangle +\sqrt{1-p} e^{i\phi }|1\rangle
$, with $\langle 1|0\rangle =0$. On the other hand, we will show that, quite surprisingly, this
no-go theorem does not hold for continuous variables.  In fact, we will show that we can make
uncorrelated cloning with a slight modification of the customary setup for cloning coherent states.

The paper is organized as follows. In Sec. II we review the general problem of optimal state
decorrelation. In Sec. III we show the general structure of the quantum channels that erase
correlations for covariant set of states, when both different and identical signals are encoded on
the local states of a multipartite density matrix. In Sec. IV the theory is specialized to the case
of two qubits with detailed derivation of the results, and the special form of the set of
decorrelable states is obtained. In Sec. V we give the proof that approximate cloning without
correlations for continuous sets of qudit states is impossible. The case of continuous variables is
reviewed in Sec. VI, where we show that an arbitrary set of bipartite Gaussian state can be
decorrelated in a covariant way with respect to group of displacement operators, i.e. independently
of the coherent signal. Moreover, we show that it is possible to realize continuous variable cloning
without correlation between the copies.  Sec. VII is devoted to the conclusions and discussion of
open problems.

\section{The problem of optimal decorrelation}
We say that a quantum channel $\map{D}$ \emph{decorrelates exactly} an $N$-partite state $\rho$
if the following equation holds:
\begin{equation}
\label{eq:decorgeneral} \map{D}\left(\rho \right) =
[\rho]_1 \otimes\dots\otimes[\rho]_N,
\end{equation}
where $[\rho]_i$ is the local state of the $i$-th party, which is given by the reduced density matrix
of $\rho$
\begin{equation}
[\rho]_i:=\Tr_{1,\ldots,i-1,i+1,\ldots,N}[\rho]
\end{equation}
The problem of state-decorrelability is the following: given a set of states $\sS$, we ask whether there
exists a quantum channel $\map{D}$ that satisfies (\ref{eq:decorgeneral}) for every state
$\rho\in\sS$.  As for the no-cloning theorem, the answer will strongly depend on the set of states
$\sS$. In particular, if the set $\sS$ consists of only one element $\rho$, then the problem of
decorrelability is trivial (one considers the channel producing $\otimes _{i=1}^N [\rho]_{i}$ for all
input states). On the other hand, if $\sS$ is the set of all possible density matrices,
decorrelation is forbidden by linearity of quantum mechanics \cite{Terno}. A stronger conclusion
immediately follows \cite{our1}: if $\sS$ contains the states $\rho'$, $\rho''$ and their convex
combination $\lambda \rho' +(1-\lambda) \rho''$, and $\rho'$ and $\rho''$ differ at least on two
parties, then exact decorrelability of $\sS$ is impossible. Impossibility of exact decorrelability
of some two-state sets can be proved \cite{Terno} due to increase in state distinguishability (see
also \cite{Mor} for some results on disentangling rather than decorrelating states).  Notice,
however, that non-increase in distinguishability of states is a necessary, but not a sufficient
condition for decorrelability.

The approximate state-decorrelation problem that we want to address here is the decorrelation of an
unknown state while preserving as much as possible the features of the local states.  More
precisely, with an information-theoretical motivation, as in Ref. \cite{our1} we will consider the
following problem \medskip\par {\bf Problem: [Optimally locally-faithful decorrelation of symmetric
  sets of states]}.  {\em Consider a set of states of the form
\begin{equation}\label{orb}
  \sS=\{\rho_{\vec g}\},\;
  \rho_{\vec g}:=U_{\vec g}\rho_{\vec e} U_{\vec g}^\dag,\quad
  U_{\vec g}:=U_{g_1}\otimes\dots\otimes U_{g_N}.
\end{equation}
where $\gG$ is a group, $U_g$ $g\in\gG$ are unitary operators acting over the Hilbert space $\sH$ of
the local quantum system, ${\vec g}=(g_1, \dots, g_N)\in\gG^N$, and the ``seed'' state $\rho_{\vec e}$ is an $N$-partite \emph{correlated} state. Find a channel $\map{D}$ that decorrelates all
states in $\sS$, namely
\begin{equation}
\label{eq:decorseed}
\forall\rho\in\sS,\;\map{D}(\rho)=[\tilde{\rho}]_1 \otimes \dots [\tilde{\rho}]_N,
\end{equation}
where $[\tilde{\rho}]_i$ is not necessary equal to the local state $[\rho]_i$, and is optimally locally
faithful, i.~e. it maximizes the averaged local fidelity
\begin{equation}\label{eq:fidel}
  \overline F[\rho_{\vec e},\map{D}]=
\frac1N\sum_{i=1}^N\int_{\gG^N}\!\!\!\!\!d{\vec g}\ F([\rho_{\vec g}]_i,[\map{D}(\rho_{\vec g})]_i),
\end{equation}
where $d{\vec g}$ denotes the Haar measure of the group \cite{note:Haar}.}

As a result of the application of channel $\map{D}$, subsystems become perfectly decorrelated,
however, at expense of losing some information about local states.  The faithfulness of
decorrelation will be judged based on the fidelity between input and output local states, averaged
over systems and over the group.  The seed state $\rho_{\vec e}$ (${\vec e}$ denotes the identity
element of $\gG^N$) plays the role of the noisy carrier on which the ``signals'' $\vec g\in\gG^N$
are encoded by the unitary modulation $U_{\vec g}=U_{g_1}\otimes\ldots\otimes U_{g_N}$.  The unitary
operators, being local, do not affect the correlation of the seed state, whence all states of the
set have the same correlation.  The problem of decorrelation is now to find a channel $\map{D}$ that
decorrelates all states of the form (\ref{orb}) while optimally preserving the signal on local
states.  The word ``signal'' may suggest a sequence of pieces of information being transmitted: in
our case this will correspond to sequels of preparations of states within the ensemble described by
$\rho_{\vec g}$. We emphasize that in the present framework we are not dealing with decorrelation of
\emph{signals}, but rather with decorrelation of states carrying them. Hence, there is no
contradiction in performing decorrelation and still claiming, e.g., that the encoded signals are
identical, e.~g. when $g_1=\ldots=g_N$.

The figure of merit (\ref{eq:fidel}) is a natural choice, in consideration of the special form
(\ref{orb}) of the set $\sS$ to be decorrelated as orbit of the seed state $\rho$ under the group
$\gG^N$. Using the fact that $\sqrt{U_{\vec g}\rho U_{\vec g}^\dag}=U_{\vec g}\sqrt\rho U_{\vec
  g}^\dag$, along with the strong concavity of the Uhlmann fidelity, we obtain the following bound
\begin{equation}
\overline F[\rho_{\vec e},\map{D}]\leq\frac1N\sum_{i=1}^NF\left([\rho_{\vec e}]_i,\int_{G^N}d{\vec g}\
    U_{g_i}^\dag[\map{D}(\rho_{\vec g})]_iU_{g_i}\right).
\end{equation}
From the last inequality it is clear that the group-averaged map
\begin{equation}\label{covar}
  \tilde{\map{D}}(\rho)=\int_{G^N}d{\vec g}\ U_{\vec g}^\dag \map{D}(U_{\vec g}\rho U_{\vec g}^\dag)U_{\vec g}
\end{equation}
has always greater or equal than that achieved by $\map{D}$. The map $\tilde{\map{D}}$ is
covariant under the group $\gG^N$ (shortly $\gG^N$-covariant), i.~e. for all states $\rho$ it
satisfies the identity
\begin{equation}\label{eq:covariance}
\tilde{\map{D}}(U_{\vec g}\rho U_{\vec g}^\dag)=U_{\vec g}\tilde{\map{D}}(\rho) U_{\vec g}^\dag.
\end{equation}
Since every $\gG^N$-covariant map is the group-average of itself, we can restrict the search of the
optimal map to covariant maps only.

Notice that for a covariant channel $\map{D}$ it is sufficient to decorrelate only one state of
$\sS$, since then it will automatically decorrelate all states of the set.  Therefore, the problem
is reduced to find a $\gG^N$-covariant map that decorrelates only the seed state (notice that,
however, this does not trivialize the problem, since the channel that sends all states to the same
fixed decorrelated state is not covariant).

If we have additional constraints on the signals (e.~g.  we know that they are identical) the set
$\sS$ becomes smaller and the problem of decorrelation easier. We will also consider this special
case of tensor representation $U_g=U_g^{\otimes N}$ of the group $\gG$, i.~e. with all identical
signals $g_1=\ldots=g_N$.

In conclusion of this section we want to comment more about the fidelity figure of merit for the
case of qubits. Here the fidelity of two states has a simple expression in terms of their Bloch
vector. It is not clear, a priori, whether it is possible to have a decorrelating covariant map that
increases the length of Bloch vectors of local states (thus decreasing the fidelity).  However, as a
result of maximizing the fidelity it turns out that the Bloch vector is always shrunk, whence the
optimal fidelity corresponds to maximum length of the output local Bloch vector.  This optimization
will be carried out in detail in the next sections.

\section{Covariance constraints}
For the same reason that led us to consider only covariant decorrelation channels, we can take the
channel as permutationally covariant, namely for every $N$ party state $\rho$ we have
\begin{equation}
  \map{D}(\Pi \rho \Pi^\dagger)=\Pi \map{D}(\rho) \Pi^\dagger\;,
\end{equation}
where $\Pi$ is an arbitrary permutation of subsystems.  In the particular case in which
$g_1=\dots=g_N$, all the signals are equal and we will consider permutationally invariant input
states $\rho$. Correspondingly, we can impose a stronger permutational simmetry on the map, namely
permutational invariance both at the input and at the output, namely
\begin{equation}
  \map{D}(\Pi \rho \Pi^\dagger)=\map{D}(\rho),\quad  \Pi \map{D}(\rho) \Pi^\dagger=\map{D}(\rho)\;.
\end{equation}

\subsection{Structure of covariant channels}
Covariance constraints are conveniently expressed using Choi-Jamio{\l}kowski isomorphism.  Under
this isomorphism a completely positive map $\map{D}$ from $\Bndd{\sH^\t{in}}$ to
$\Bndd{\sH^\t{out}}$ is mapped in a one-to-one way to the positive operator $R_\map{D}\in
\Bndd{\sH^\t{out}\otimes\sH^\t{in}}$:
\begin{equation}
R_\map{D}=\Choi{D}:=\map{D} \otimes\map{I}(\ket{\Psi}\bra{\Psi}),
\end{equation}
where $\ket{\Psi}=\sum_i \ket{i}\otimes \ket{i}$ is a maximally
entangled vector in $\sH^\t{in}\otimes \sH^\t{in}$.
The trace-preserving condition of $\map{D}$ implies that
\begin{equation}
\label{eq:jamioltrace}
\Tr_{\t{out}}(R_\map{D})=\openone_{\t{in}}.
\end{equation}
One can express the state transformation using operator $R_\map{D}$ with
\begin{equation}
\map{D}(\rho)=\Tr_{\t{in}}\left(R_\map{D}\  \openone_{\t{out}}\otimes \rho^T \right).
\end{equation}
The general covariance condition
\begin{equation}
 \map{D}\left(V_g \rho V_g^\dagger \right) = W_g \map{D}(\rho) W_g^\dagger,
\end{equation}
with $V_g$ and $W_g$ unitary representations of a group, translates to the commutation condition for
$R_\map{D}$
\begin{equation}
\label{eq:covjamiolgen}
[R_\map{D},W_g\otimes V_g]=0.
\end{equation}

\subsection{Different signals}
\label{sec:covdiffsig}
Let us consider a covariant operation $\map D$ acting on $N$
qubit states fulfilling the covariance condition
(\ref{eq:covariance}), where $g_i \in SU(2)$, $U_{g}$ is the defining
representation of $SU(2)$ and we do not impose any additional
constraints on $g_i$.  The covariance condition
(\ref{eq:covjamiolgen}) applied to this case reads:
\begin{equation} [R_\map D,\underbrace{U_{g_1}\otimes\dots\otimes
    U_{g_N}}_{\mathcal{H}^\t{out}} \otimes
  \underbrace{U_{g_1}^*\otimes\dots\otimes
    U_{g_N}^*}_{\mathcal{H}^\t{in}}]=0.
\end{equation}
Since for $SU(2)$ group the conjugated representation $U_g^*$ is
equivalent to $U_g$, we may simplify the above condition by
introducing the new operator
\begin{equation}
  \overline{R}_\map D=\openone_{\mathcal{H}^\t{out}} \otimes \sigma_y^{\otimes N} \ R_\map D \ \openone_{\mathcal{H}^\t{out}} \otimes \sigma_y^{\otimes N}.
\end{equation}
For this operator the covariance condition no longer involves
conjugated representations:
\begin{equation}
  \label{eq:covjamiolconj}
  [\overline{R}_\map D,\underbrace{U_{g_1}\otimes\dots\otimes U_{g_N}}_{\mathcal{H}^\t{out}} \otimes \underbrace{U_{g_1}\otimes\dots\otimes U_{g_N}}_{\mathcal{H}^\t{in}}]=0.
\end{equation}
Evolution of the state can be expressed using
$\overline{R}_\map D$ as follows
\begin{equation}
  \map D(\rho)=\Tr_{\mathcal{H}^\t{in}}[ \overline{R}_\map D \ (\openone_{\mathcal{H}^\t{out}} \otimes
  \overline{\rho} )],
\end{equation}
where $\overline{\rho}=\sigma_y^{\otimes N} \rho^T \sigma_y^{\otimes
  N}$.  We will write the operator $\overline{R}_\map D$ by
changing the order of the Hilbert spaces, such that input and output spaces of the $i$-th qubit stand
next to each other, namely
\begin{multline}
  \mathcal{H}_1^\t{out} \otimes \dots \otimes \mathcal{H}_N^\t{out}
  \otimes \mathcal{H}_1^\t{in} \otimes \dots \otimes
  \mathcal{H}_N^\t{in}
  \rightarrow \\
  \mathcal{H}_1^\t{out} \otimes \mathcal{H}_1^\t{in} \otimes \dots
  \otimes \mathcal{H}_N^{\t{out}} \otimes \mathcal{H}_N^\t{in}.
\end{multline}
After this rearrangement the covariance condition takes the form
\begin{equation}
[\overline{R}_\map D, U_{g_1}\otimes U_{g_1}
  \dots\otimes U_{g_N} \otimes U_{g_N}]=0,
\end{equation}
which implies that $\overline{R}_\map D$ can be expressed in a
simple way using projections on two-qubit singlet ($P^{(0)}$) and
triplet ($P^{(1)}$) subspaces:
\begin{equation}
\label{eq:paramdiff}
\overline{R}_\map D=\sum_{i_1,\dots,i_N=0}^1 a_{i_1,\dots,i_N} P^{(i_1)}\otimes \dots \otimes P^{(i_N)},
\end{equation}
where $a_{i_1,\dots,i_N}$ are positive coefficients. Additionally, in
order to assure permutational covariance of $\map D$,
coefficients $a_{i_1,\dots,i_N}$ cannot depend on the order of
indices. Then, we can introduce a smaller number of coefficients
$q_n :=
a_{i_1,\dots,i_N}$, where $n$ is the number of indices $i_k$ equal to
one. The most general covariant map is thus characterized by $N+1$
nonnegative coefficients $q_n$.
Eq.~\eqref{eq:paramdiff} becomes then
\begin{equation}
  \overline{R}_\map D=\sum_{n=0}^Nq_n\left\{\sum_{\pi\in D_n} \pi \left(P^{(1)\otimes n} \otimes P^{(0)\otimes N-n}\right) \pi\right\},
\end{equation}
where $D_n$ is the set of permutation operators $\pi$ of the $N$
qubits that do not leave $P^{(1)\otimes n} \otimes P^{(0)\otimes N-n}$
invariant.  Clearly, the cardinality of $D_n$ is $\binom{N}{n}$. Since
one has $\Tr_{\mathcal{H}_i^\t{out}}[P^{(0)}]=\frac12\openone$ and
$\Tr_{\mathcal{H}_i^\t{out}}[P^{(1)}]=\frac32\openone$ for $1\leq i\leq N$,
the trace-preserving condition (\ref{eq:jamioltrace}) leads then to
the following constraint on the coefficients $q_n$
\begin{equation}
\label{eq:constrdiff}
\sum_{n=0}^N \frac{3^n}{2^N}\binom{N}{n}q_n=1.
\end{equation}
Eventually, we have $N$ independent coefficients characterizing
covariant transformations. This is the freedom that we have when attempting
to decorrelate set of states (\ref{orb}) in a covariant way in the case of
different $SU(2)$ signals being encoded. Notice that the above
characterization may be simply generalized from qubits to arbitrary
$d$ dimensional systems, by encoding signals via $SU(d)$ defining
representation (we do not use the equivalence of $U$ and $U^*$, and
$P^{(0)}=\frac1d\ket\Psi\bra\Psi$ and $P^{(1)}=\openone-P^{(0)}$).

\subsection{Identical signals}
\label{sec:covident}
We now characterize covariant operations in the case of identical
signals $g_1=\dots=g_N$. This is an especially interesting case due to
its relevance for quantum cloning, broadcasting and state estimation
problems. In this case, the  information about the quantum state
(playing the role of the signal) is distributed to many subsystems. The covariance
condition (\ref{eq:covjamiolconj}) for the $N$ qubit transformation in
the case of identical signals has form
\begin{equation}
[\overline{R}_\map D,\underbrace{U_{g}^{\otimes N}}_{\mathcal{H}^\t{out}} \otimes \underbrace{U_{g}^{\otimes N}}_{\mathcal{H}^\t{in}}]=0.
\end{equation}
This is a much weaker condition than (\ref{eq:covjamiolconj}),  and hence
the structure of covariant operations will be significantly reacher.
Recall that an $N$-fold tensor product of two-dimensional Hilbert spaces can
be decomposed with respect to the action of $U^{\otimes N}$ in the
following way
\begin{equation}
  \mathcal{H}^{\otimes N}=\bigoplus_{j=s_N}^{N/2} \mathcal{H}_j \otimes \mathbb{C}^{\kappa_j},
\end{equation}
where $s_N=(N\ \t{mod}\ 2)/2$, $\mathcal{H}_j$ carries an irreducible
representation of $SU(2)$ corresponding to the total angular momentum
$j$, and
\begin{equation}
  \kappa_j=\frac{2j+1}{N/2+j+1}\binom{N}{N/2+j}
\end{equation}
denotes the multiplicity of this representation. To evaluate the
operator $\overline{R}_\map D$ we will decompose the output and
input subspaces as follows
\begin{equation}
  \underbrace{\mathcal{H}^{\otimes N}}_{\mathcal{H}^\t{out}} \otimes \underbrace{\mathcal{H}^{\otimes N}}_{\mathcal{H}^\t{in}}=
  \bigoplus_{j,l=s_N}^{N/2} \mathcal{H}_j^\t{out} \otimes \mathbb{C}^{\kappa_j}_\t{out}  \otimes \mathcal{H}_{l}^\t{in} \otimes \mathbb{C}^{\kappa_l}_\t{in}.
\end{equation}
Conveniently, we change the notation order, so that the subspaces are
ordered as $\mathcal{H}_j^\t{out} \otimes \mathcal{H}_{l}^\t{in}
\otimes \mathbb{C}^{\kappa_j}_\t{out} \otimes
\mathbb{C}^{\kappa_l}_\t{in}$, and we have $\mathcal{H}_j^\t{out}
\otimes \mathcal{H}_{l}^\t{in} = \bigoplus_{J=|j-l|}^{j+l}
\mathcal{H}_J$. We will focus now attention to the simple case of
permutationally invariant seed state, and hence permutationally
invariant output state. Therefore, without loss of
generality, we can limit the optimization to maps with permutationally
invariant input and output. It turns out that the irreducible
spaces for the permutations of $N$ systems are exactly the
multiplicity spaces $\mathbb{C}^{\kappa_j}$ for the irreducible
representations of $U^{\otimes N}$. This implies that permutational
invariance selects maps of the form
\begin{equation}
  \overline{R}_{\map D}=\bigoplus_{j,l=s_N}^{N/2}R^{(jl)}\otimes\openone_{\kappa_j} \otimes \openone_{\kappa_l}.
\end{equation}
Finally, it can be easily shown that the covariance condition above
together with the permutational invariance leads to the following
structure of the operator $\overline{R}_\map D$ \cite{our}
\begin{equation}
  \overline{R}_\map D=\bigoplus_{j,l=s_N}^{N/2}\bigoplus_{J=|j-l|}^{j+l} s_{j,l}^J P_{j,l}^{(J)} \otimes \openone_{\kappa_j} \otimes \openone_{\kappa_l},
\end{equation}
where $s_{j,l}^J$ are nonnegative coefficients and $P_{j,l}^{(J)} \in
\Bndd{(\mathcal{H}_j^\t{out} \otimes \mathcal{H}_l^\t{in}}$ is a
projector on the subspace $\mathcal{H}_J$ with total angular momentum
$J$. The trace-preserving condition is given by \cite{our}
\begin{equation}
\label{eq:constrprid}
\sum_{j=s_N}^{N/2} \sum_{J=|j-l|}^{j+l} \frac{2J+1}{2l+1}\kappa_j
s_{j,l}^J = 1,\ \  \forall
l:\ s_N \leq l \leq \frac{N}{2}.
\end{equation}
Up to the leading order in $N$, the number of independent parameters
$s_{j,l}^{J}$ scales as $N^3/6$, which  reflects the fact that
covariance condition in the case of identical signals is much weaker
than in the case of different ones, where the leading order of the
scaling is $N$.

\section{Decorrelability of qubits}
The problem of decorrelability of $N$ qubit states can now be stated
in a simple way. Without loss of generality we may assume that the
single qubit reduced density matrices of the seed state $\rho$ are
diagonal in the $\sigma_z$ eigenbasis, i.e. have the form
$\rho_i=1/2(\openone + \eta \sigma_z)$, where $\eta$ is the length of
the Bloch vector.  The set of $N$ qubit states is nontrivially
decorrelable in the different (identical) signal scenario if there
exist positive parameters $q_n$ ($s_{j,l}^J$) satisfying the
trace-preserving constraints in Eq.~(\ref{eq:constrdiff})
(Eq.~(\ref{eq:constrprid})), such that the corresponding map generates
a product state from the seed $\rho $, namely
\begin{equation}
\label{eq:decorexplicit}
\Tr_{\mathcal{H}^\t{in}}[\overline{R}_\map D\ (\openone_{\mathcal{H}^\t{out}} \otimes
\overline{\rho} )]= [1/2(\openone+\tilde{\eta} \sigma_z)]^{\otimes N},
\end{equation}
with $\tilde{\eta}>0$ ($\tilde{\eta}=0$ would mean a complete loss of information). The maximum achievable $\tilde{\eta}$
is a measure of quality of decorrelation process. The interesting
question is now for which kind of seed states decorrelation is
possible and for which kind of seed
states it is not.

We now present the full solution for the simplest case of two qubits.
Consider a couple of qubits $A$ and $B$. Permutational invariance of
the seed state $\rho_{AB}$, along with the condition that local states
are diagonal in the $\sigma_z$ eigenbasis implies that $\rho_{AB}$ has
the form
\begin{equation}
  \rho_{AB}=\frac14\left(\openone+\eta(\sigma_z\otimes \openone+\openone\otimes\sigma_z)-\sum_{i,j=x,y,z}\lambda_{ij}\sigma_i\otimes\sigma_j\right),
\label{eq:perminv}
\end{equation}
with $\lambda_{ij}=\lambda_{ji}$.

\subsection{Different signals}
\begin{figure}[ht]
\includegraphics[width= 0.5\textwidth]{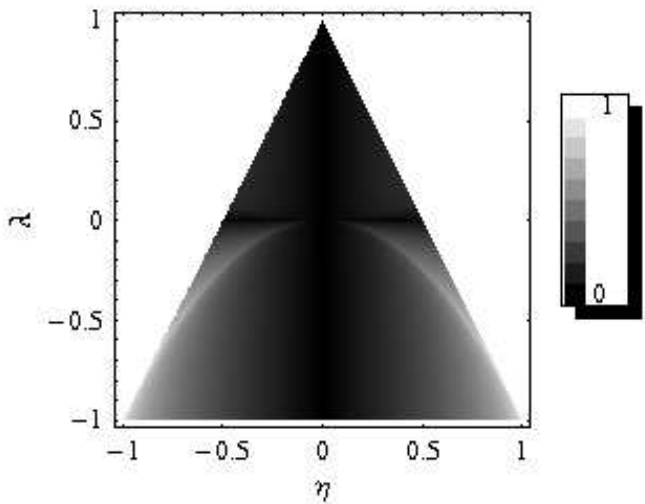}
\caption{Length $\tilde{\eta} $ of the Bloch vectors of the decorrelated
  states of two qubits starting from the joint state in Eq.~(\ref{eq:pauli}). The plot depicts the maximal achievable
  $\tilde{\eta} $ in gray scale versus the parameters $\eta$ and $\lambda$ of the
  input state.}
\label{f:eta}
\end{figure}

Applying the general results of Sec.~\ref{sec:covdiffsig}, we find that a covariant operation $\map D$ is parameterized with three parameters $q_0$, $q_1$, $q_2$
[see Eq.~(\ref{eq:paramdiff})] satisfying the trace-preserving condition
\begin{equation}
\label{eq:trprdiff2}
q_0 + 6q_1+9q_2=4\;,
\end{equation}
and one has
\begin{equation}
\begin{split}
  \overline{R}_\map D=&q_0 P^{(0)}\otimes P^{(0)} \\+& q_1
  (P^{(0)}\otimes P^{(1)}+ P^{(1)}\otimes P^{(0)})\\ +&q_2
  P^{(1)}\otimes P^{(1)}.
\end{split}
\end{equation}
In order to get a better intuition, we write explicitly the map as
follows
\begin{equation}
  \map D(\rho_{AB})=\frac{q_0}{4}  \map D_0(\rho_{AB}) + \frac{3 q_1}{2} \map D_1(\rho_{AB}) + \frac{9 q_2}{4} \map D_2(\rho_{AB}),
\end{equation}
where $\map D_i$ are the trace-preserving maps
\begin{align}
  \map D_0(\rho_{AB})=&\rho_{AB}\;,\\
  \map D_1(\rho_{AB})=&\tfrac{1}{3}\left(
    \rho_A \otimes \openone + \openone \otimes \rho_B - \rho_{AB} \right) \;,\\
  \map D_2(\rho_{AB})=&\tfrac{1}{9}\left(4 \openone \otimes
    \openone -2 \rho_A \otimes \openone- 2 \openone \otimes \rho_B +
    \rho_{AB}\right) \;.
\end{align}
Using the decorrelability condition (\ref{eq:decorexplicit}) and the
expression of $\rho_{AB}$ in Eq.~\eqref{eq:perminv} we obtain that
decorrelation is possible when $\lambda_{ij}=0$ apart from
$\lambda_{zz}:=\lambda$. Decorrelation then corresponds to the
following conditions
\begin{align}
  q_0&=\frac{1}{4}\left(1 +\frac{6 \tilde{\eta}}{\eta}- \frac{9
      \tilde{\eta}^2}{\lambda} \right) \;,\label{eq:diageq1}\\
  q_1&=\frac{1}{4}\left(1+\frac{2 \tilde{\eta}}{\eta}+ \frac{3
      \tilde{\eta}^2}{\lambda}\right) \;,\\
  q_2&=\frac{1}{4}\left(1-\frac{2 \tilde{\eta}}{\eta}-
    \frac{\tilde{\eta}^2}{\lambda}\right).
\label{eq:diageq3}
\end{align}
Analysis of the above equations (together with the trace preserving
condition (\ref{eq:trprdiff2})) leads to the following conclusions.
Equations are always satisfied for arbitrary seed state $\rho$ for:
$q_0=1/4$, $q_1=1/4 $, $q_2=1/4$. This case is, however, not of much
interest since it corresponds to a completely mixing channel resulting
in $\tilde{\eta}=0$, and hence destroying all encoded information.
We can now write decorrelable states as in Eq.~\eqref{eq:perminv}
\begin{equation}
\label{eq:pauli}
\rho_{AB}=\frac{1}{4}\left[\openone \otimes \openone + \eta
  (\sigma_z \otimes \openone + \openone \otimes \sigma_z) - \lambda
  \sigma_z \otimes \sigma_z \right],
\end{equation}
where positivity corresponds to the following conditions
\begin{equation}
  |\eta|\leq1,\quad\lambda\leq1-|\eta|.
\end{equation}
Notice that all states of the form (\ref{eq:pauli}) are separable, by
just using the PPT criterion \cite{ppt}. Finally, to get the optimal
decorrelation qualityfind which states are decorrelable we find
solutions of Eqs.~(\ref{eq:diageq1}-\ref{eq:diageq3}) with the
maximally achievable $\tilde{\eta}$, which is
\begin{align}
  & |\tilde{\eta}| = \frac{-\lambda -\sqrt{\eta^2\lambda+\lambda^2}}{|\eta|} &  \quad -1 \leq \lambda \leq -\eta^2\\
  & |\tilde{\eta}| = \frac{-\lambda +\sqrt{\lambda^2 - 3\eta^2\lambda}}{3|\eta|} & \quad -\eta^2 \leq \lambda \leq 0\\
  & |\tilde{\eta}| = \frac{\lambda + \sqrt{\eta^2\lambda+\lambda^2}}{3|\eta|} & \quad 0 \leq \lambda \leq \eta^2/3\\
  & |\tilde{\eta}| = \frac{-\lambda +
    \sqrt{\eta^2\lambda+\lambda^2}}{|\eta|} & \quad \eta^2/3 \leq
  \lambda \leq 1.
\end{align}
This solution is plotted in Fig.~\ref{f:eta}.
The visible parabola in the picture corresponds to the initial states which are already in the product form, i.e.
\begin{multline}
\rho_{AB}=[1/2\left(\openone + \eta \sigma_z \right)]^{\otimes 2} = \\\frac{1}{4} \left[
\openone\otimes \openone + \eta (\sigma_z \otimes \openone + \openone \otimes \sigma_z) + \eta^2
\sigma_z \otimes \sigma_z   \right]\;.
\end{multline}
Clearly, such states are trivially decorrelable, as they are already
decorrelated, and $\eta^\prime=\eta$. These states correspond to the
case $\lambda=-\eta^2$, and this explains the parabolic
structure in the figure.

\subsection{Identical signals}
\label{sec:decorident}
We introduce the following notation
to  denote bipartite vectors
\begin{equation}
  \kket{A}:=\sum_{m,n=0}^1A_{mn}\ket m\otimes \ket n,
\label{dket}
\end{equation}
where $A_{mn}$ are the matrix elements on the basis $\{\ket i\}$ of
the operator $A$. The useful properties of this notation are the
following
\begin{equation}
  A\otimes C\kket B=\kket{ABC^T},\quad\bbrakket{A}{B}=\Tr[A^\dagger B].
\end{equation}
Consider the situation in which signals encoded on the
two qubits are equal. In particular, this is the situation
after performing $1\to 2$ universal cloning of qubits, starting form
an unknown input state $\ket{\psi}=U\ket{0}$. The optimal cloning
operation produces two clones in the state: $U\otimes U \rho_{AB}
U^\dagger \otimes U^\dagger$, where
$\rho_{AB}=2/3\ket{00}\bra{00}+1/3\ket{\psi^+}\bra{\psi^+}$ and
$\ket{\psi^+}=1/\sqrt{2}\kket{\sigma_x}$. Notice that this is a
correlated state.

We want to know which two-qubit states are decorrelable and what is
the maximal attainable length of the output Bloch vector
$\tilde{\eta}$.  Since we now impose a weaker covariance condition, we
expect decorrelation to succeed for a larger class of states than in
the case of independent signals.

Using the general covariance conditions described in Sec.~\ref{sec:covident}, we
get a parametrization of covariant operations using six
parameters $s_{j,l}^J$, that for convenience we relabel as: $q_0=s_{0,0}^0$,
$q_1=s_{1,0}^1$, $q_2=s_{0,1}^1$, $q_3=s_{1,1}^0$, $q_4=s_{1,1}^1$,
$q_5=s_{1,1}^2$.  The trace-preserving conditions are rewritten as follows
(\ref{eq:constrprid}) read
\begin{equation}
\label{eq:trprid2}
q_0+3q_1=1,  \quad q_2 + \frac{1}{3}q_3 +q_4 +\frac{5}{3}q_5 =1.
\end{equation}
The projections $P^J_{j,l}$ can be written as follows using the
notation of Eq.~\eqref{dket}
\begin{align}
  P^{(0)}_{00}&=\frac14\kket{\sigma_y}\bbra{\sigma_y}\otimes\kket{\sigma_y}\bbra{\sigma_y},\\
  P^{(1)}_{10}&=\frac12\left(\openone-\frac12\kket{\sigma_y}\bbra{\sigma_y}\right)\otimes\kket{\sigma_y}\bbra{\sigma_y},\\
  P^{(1)}_{01}&=\frac12\kket{\sigma_y}\bbra{\sigma_y}\otimes\left(\openone-\frac12\kket{\sigma_y}\bbra{\sigma_y}\right),\\
  P^{(0)}_{11}&=\frac1{12}\sum_{i=0,x,z}s(i)\kket{\sigma_i}\kket{\sigma_i}\sum_{j=0,x,z}s(j)\bbra{\sigma_j}\bbra{\sigma_j},\\
  P^{(1)}_{11}&=\frac1{16}\sum_{i,j=0,x,z}(\kket{\sigma_i}\kket{\sigma_j}-\kket{\sigma_j}\kket{\sigma_i})\\
  &\times(\bbra{\sigma_i}\bbra{\sigma_j}-\bbra{\sigma_j}\bbra{\sigma_i}),\\
  P^{(2)}_{11}&=\frac1{16}\sum_{i,j=0,x,z}(\kket{\sigma_i}\kket{\sigma_j}+\kket{\sigma_j}\kket{\sigma_i})\\
  &\times(\bbra{\sigma_i}\bbra{\sigma_j}+\bbra{\sigma_j}\bbra{\sigma_i})-P^0_{11},
\end{align}
where $\sigma_0:=\openone$ and $s(0)=1$ $s(x)=s(z)=-1$. The action on
the states $\rho_{AB}$ of Eq.~\eqref{eq:perminv} of the (normalized)
maps corresponding to each of the operators above is the following
\begin{align}
  \map D^{(0)}_{00}(\rho_{AB})&=P^{(0)}\Tr[P^{(0)}\rho_{AB}],\\
  \map D^{(1)}_{10}(\rho_{AB})&=\frac13P^{(1)}\Tr[P^{(0)}\rho_{AB}],\\
  \map D^{(1)}_{01}(\rho_{AB})&=P^{(0)}\Tr[P^{(1)}\rho_{AB}],\\
  \map D^{(0)}_{11}(\rho_{AB})&=P^{(1)}\rho_{AB}P^{(1)},\\
  \map D^{(1)}_{11}(\rho_{AB})&=\frac12(P^{(1)}\Tr[P^{(1)}\rho_{AB}]-P^{(1)}\overline\rho_{AB}P^{(1)}),\\
  \map D^{(2)}_{11}(\rho_{AB})&=\frac3{10}(P^{(1)}\Tr[P^{(1)}\rho_{AB}]+
  P^{(1)}\overline\rho_{AB}P^{(1)})\\
  &-\frac15P^{(1)}\rho_{AB}P^{(1)},
\end{align}
The most general covariant and permutation invariant map is then of
the form
\begin{equation}
\begin{split}
  \map D(\rho_{AB})&=q_0\map D^0_{00}(\rho_{AB})+3q_1\map D^1_{10}(\rho_{AB})+q_2\map D^1_{01}(\rho_{AB})\\
  &+\frac{q_3}3\map D^0_{11}(\rho_{AB})+q_4\map D^1_{11}(\rho_{AB})+\frac{5q_5}3\map D^2_{11}(\rho_{AB}),
\end{split}
\end{equation}
We can now write the output state as follows
\begin{align}\label{covmap}
  {\map D}(\rho_{AB})=&\left(\frac{q_3}{3}-\frac{q_4}2+\frac{q_5}{6}\right)\rho_{AB}\nonumber\\
  &+\left(q_4-q_5\right)\left(\frac{\eta}4(\sigma_z\otimes \openone+\openone\otimes\sigma_z)\right)\nonumber\\
  &+\frac14\left(q_0-\frac{q_3}3+\frac{q_4}2-\frac{q_5}6\right)(1+\Lambda)P^{(0)}\nonumber\\
  &+\frac14\left(\frac{q_4+q_5}2\right)(3-\Lambda)P^{(1)}\nonumber\\
  &+\frac{q_2}{4}(3-\Lambda)P^{(0)}+\frac{q_1}4(1+\Lambda)P^{(1)},
\end{align}
where $\Lambda=\lambda_{xx}+\lambda_{yy}+\lambda_{zz}$. If we consider
the terms in $\sigma_i\otimes\sigma_j$ with $i\neq j$, it is clear
that either $\frac{q_3}{3}-\frac{q_4}2+\frac{q_5}{6}=0$, or it is
impossible to decorrelate the input state. However, the condition
$\frac{q_3}{3}-\frac{q_4}2+\frac{q_5}{6}=0$ would lead to trivial
decorrelation, with total loss of information. We must then have
$\lambda_{ij}=0$ for $i\neq j$ at the input state. Moreover,
considering that
\begin{equation}
  P^{(0)}=\frac14(\openone\otimes\openone-\sigma_x\otimes\sigma_x-\sigma_y\otimes\sigma_y-\sigma_z\otimes\sigma_z),
\end{equation}
and $P^{(1)}=\openone-P^{(0)}$, the only term in Eq.~\eqref{covmap}
containing $\sigma_i^{\otimes 2}$ with possibly different weights is
the first one, in order to have ${\map D}(\rho_{AB})$ without
terms in $\sigma_x^{\otimes2}$ or in $\sigma_y^{\otimes2}$, we must
have $\lambda_{xx}=\lambda_{yy}$, namely decorrelable states are of
the form
\begin{equation}
\begin{split}
  \rho_{AB}&=\frac{1}{4}[\openone\otimes \openone +
  \eta (\sigma_z \otimes \openone + \openone \otimes \sigma_z)\\
  &-\frac{\Lambda-\lambda}2(\sigma_x \otimes \sigma_x +\sigma_y \otimes
  \sigma_y)  - \lambda \sigma_z \otimes \sigma_z].
\end{split}
\end{equation}

\subsubsection{Symmetric Input state}
Let us first restrict to seed states supported on symmetric subspace,
i.e. $\Tr[P^{(0)}\rho_{AB}]=\frac14(1+\Lambda)=0$ (this set of states
contains the states produced by $1\to 2$ optimal universal cloning
machine). The relevant variables in this case are $q_3$, $q_4$ and
$q_5$, since $q_0$ and $q_1$ do not enter the equations and $q_2$ is
automatically determined by $q_2=(1-\tilde{\eta}^2)/4$.

\begin{figure}[ht]
   \includegraphics[width=
   0.5\textwidth]{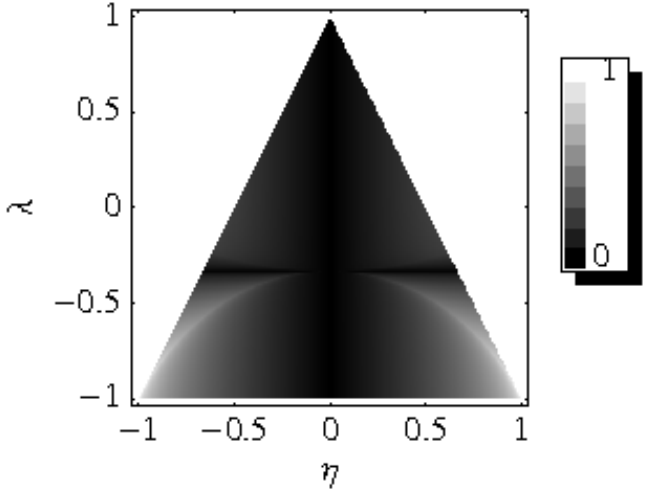}
   \caption{Length $\tilde \eta $ of the Bloch vectors of the
     decorrelated states of two qubits starting from a seed state
     supported on the symmetric subspace parameterized as in
     Eq.~(\ref{eq:paulisym}). The plot depicts the maximal achievable
     $\tilde \eta $ versus the parameters $\eta$ and $\lambda$ of the
     input state.}
 \label{f:eta2}
 \end{figure}

In terms of the variables $\eta$ (length of the initial Bloch vector
of reduced density matrix) and $\lambda$, we can write symmetric
decorrelable states states using Pauli matrices as
\begin{equation}
\begin{split}
  \label{eq:paulisym}
  \rho^{\textrm{sym}}_{AB}=\frac{1}{4}[\openone\otimes \openone +
  \eta (\sigma_z \otimes \openone + \openone \otimes \sigma_z) + \\
  (1+\lambda)/2\ (\sigma_x \otimes \sigma_x + \sigma_y \otimes
  \sigma_y ) - \lambda \sigma_z \otimes \sigma_z].
\end{split}
\end{equation}
Starting from Eqs.~(\ref{eq:decorexplicit}) and \eqref{eq:perminv}, we
find that a non trivial solution to the decorrelation problem exists
provided that $\eta\neq 0$ and $\lambda \neq -1/3$, and one has
\begin{align}
  \label{eq:detnonzeroparam1}
  q_3&=\frac{1}{12}\left[3+\tilde{\eta}\left(\tilde{\eta}-\frac{40\tilde{\eta}}{1+3\lambda}+\frac{12}{\eta}
    \right) \right],
  \\
  q_4&=\frac{1}{12}\left[3+\tilde{\eta}\left(\tilde{\eta}+\frac{20\tilde{\eta}}{1+3\lambda}+\frac{6}{\eta}
    \right) \right],
  \\
  \label{eq:detnonzeroparam3}
  q_5&=\frac{1}{12}\left[3+
    \tilde{\eta}\left(\tilde{\eta}-\frac{4\tilde{\eta}}{1+3\lambda}-\frac{6}{\eta}
    \right) \right].
\end{align}
Looking for the maximal $\tilde{\eta}$ that keeps $q_i$ nonnegative we
obtain
\begin{widetext}
  \begin{align}
\label{eq:rmaxsym1}
&\tilde{\eta}=\frac{-(1+3\lambda) - \sqrt{(1+3\lambda)^2 +
    \eta^2[1+(2-3\lambda)\lambda]}}{|\eta|(1-\lambda)}, &-1 \leq
\lambda \leq \lambda_1,
\\
&\tilde{\eta}=\frac{-(1+3\lambda) + \sqrt{(1+3\lambda)[1+3\lambda
    -\eta^2(7+\lambda)]}}{|\eta|(7+\lambda)}, & \lambda_1 \leq \lambda
\leq -\frac{1}{3},
\\
&\tilde{\eta}=\frac{2(1+3\lambda)
  +\sqrt{(1+3\lambda)[\eta^2(13-\lambda) +
    4(1+3\lambda)]}}{|\eta|(13-\lambda)}, &-\frac{1}{3} \leq \lambda
\leq \lambda_2,
\\
\label{eq:rmaxsym4}
&\tilde{\eta}=\frac{-(1+3)\lambda + \sqrt{(1+3\lambda)^2 +
    \eta^2[1+(2-3\lambda)\lambda]}}{|\eta|(1-\lambda)}, &
\lambda_2\leq \lambda \leq 1,
\end{align}
\end{widetext}
where
\begin{equation}
  \lambda_1= \frac{1}{3}\left(2\sqrt{4-3\eta^2}-5 \right),   \quad \lambda_2= \frac{1}{3}\left(7-2\sqrt{16-3\eta^2} \right).
\end{equation}
See Fig.~\ref{f:eta2} for visualisation of these results.
 It is worth observing that undecorrelable states corresponding to
 $\lambda=-1/3$ are exactly those that can be obtained by a 1-to-2
 universal cloning machine. This is a manifestation
 of a general theorem of no-cloning without correlations proved in
 Sec.~\ref{sec:cloning}.

\subsubsection{Permutationally invariant input state}
A general two-qubit state containg also a singlet fraction can be
written as:
\begin{equation}
  \rho_{AB}=p\ket{\Psi^-}\bra{\Psi^-}+(1-p)\rho^\t{sym}_{AB}.
\end{equation}
Without writing and analyzing equations which is a bit tedious we just
summarize the final results.  If either $p=1$, $\lambda=-1/3$ or
$\eta=0$, then  non-trivial decorrelation is impossible (notice that
$\lambda$ and $\eta$ are calculated from the symmetric fraction
of the state: $\rho_{AB}^\t{sym}$ in the same way as in the previous
subsection).  Otherwise, two situations may occur. (i) if $\tilde{\eta}$
evaluated by Eqs.~(\ref{eq:rmaxsym1}--\ref{eq:rmaxsym4})
fulfills the condition $1-\tilde{\eta}^2-4p \geq 0$, then this is a
valid maximal achievable length of the output Bloch vector also in the
case when the state contains a singlet fraction; (ii) otherwise
$\tilde{\eta}$ should be calculated as follows.  For $-1 \leq \lambda
\leq \lambda^\prime_1$ or $\lambda^\prime_2\leq \lambda \leq 1$
\begin{equation}
\label{eq:rmaxgen1}
\tilde{\eta}=\frac{\sqrt{\alpha}}{2|\eta| \sqrt{2}}
\sqrt{9\alpha + 8 \eta^{2}(1-p) - 3\sqrt{\alpha[9\alpha + 16\eta^{ 2}(1-p)] } };
\end{equation}
for $\lambda^\prime_1 \leq \lambda \leq -\frac{1}{3}$
\begin{equation}
  \tilde{\eta}=\frac{\sqrt{\alpha}}{10|\eta| \sqrt{2}}
  \sqrt{9\alpha-40\eta^{2}(1-p) + 3\sqrt{\alpha[9\alpha - 80\eta^{ 2}(1-p)]}};
\end{equation}
for $-\frac{1}{3} \leq \lambda \leq \lambda^\prime_2$
\begin{equation}
\label{eq:rmaxgen3}
\tilde{\eta}=\frac{\sqrt{\alpha}}{10|\eta| \sqrt{2}}
\sqrt{9\alpha+20\eta^{ 2}(1-p) + 3\sqrt{\alpha[9\alpha +40\eta^{ 2}(1-p)]}};
\end{equation}
where $\alpha= 1 + 3\lambda$ and
\begin{equation}
  \lambda^\prime_1=-\frac{1}{3}\left[1+2\eta^{2}(1-p) \right],   \  \lambda^\prime_2=-\frac{1}{3}\left[1-\eta^{2}(1-p) \right].
\end{equation}
One can summarize this by observing (which may not be evident from the
above equations) that adding a singlet fraction decreases the
achievable $\tilde{\eta}$, but otherwise does not qualitatively change
the decorrelability of states. In particular, the completely
nondecorrelable states are still those that have $\lambda=-1/3$ or
$\eta=0$ in their symmetric fraction.

\section{No approximate cloning without correlations for qudit
  continuous sets of states}
\label{sec:cloning}
In Sec.~\ref{sec:decorident} we noticed that two-qubit states obtained via universal $1\to 2 $
cloning of a single qubit cannot be decorrelated. The same statement holds for clones obtained via
phase-covariant $1 \to 2$ cloning.  More generally, here we will show that there does not exist an
approximate $N$-to-$M$ cloning channel of $d-$dimensional systems (qudits) such that the obtained
clones are decorrelated, if the cloning channel is to work at least for a \emph{phase-set} of
states. By a phase-set we mean a set containing states of the form
\begin{equation}\label{phasestat}
  \ket{\phi}:=\sqrt{p} \ket{0}+\sqrt{1-p} e^{i \phi}\ket{1},
\end{equation}
for some finite continuous range of phases
$\phi$, where $\ket{0}$, $\ket{1}$ are some orthogonal vectors and $p$
is a real number $0<p<1$.  Of course, this implies that
clones obtained from any cloning machines working for a
\emph{phase-set} of states (such as e.g.  universal, phase covariant,
etc.) cannot be decorrelated.

In order to assure the full generality of the proof, we allow cloning
to be both asymmetric, and not necessarily covariant.
Consider a channel $\Lambda$, which acting on $N$ copies of a qudit state produces $M$ ($M>N$)
approximate, possibly different clones which are required to be \emph{uncorrelated}:
\begin{equation}
\label{eq:decor}
\Lambda(\ket{\phi}\bra{\phi}^{\otimes N}) =
\bigotimes_{k=1}^M \rho_k^\phi.
\end{equation}
We will show that such a transformation is impossible, if one requires
that every clone $\rho_k^{\phi}$ carries some (possibly
infinitesimally small) information on the identity of the input state
$\ket{\phi}$ and additionally that the channel works at least for
all states from some \emph{phase-set}.

Since the channel should work for states from a phase-set, let us consider its action on states
$\ket{\phi}=\sqrt{p}\ket{0}+\sqrt{1-p}e^{i \phi} \ket{1}$. Notice that the input product state
$\ket{\phi}\bra{\phi}^{\otimes N}$ depends on the phase $\phi$ via linear functions of $e^{i n
  \phi}$, where $n \in \{-N,\dots,N\}$. Thanks to linearity of $\Lambda$, the dependence of the
output state $\Lambda(\ket{\phi}\bra{\phi}^{\otimes N})$ on $\phi$ has the same character.

Consider now a map $\Lambda_k$ which is obtained from the map
$\Lambda$ [Eq.~(\ref{eq:decor})] by tracing out all output qudits
except the qudit number $k$. Its action clearly reads:
\begin{equation}
  \Lambda_k(\ket{\phi}\bra{\phi}^{\otimes N})=\rho_k^{\phi}.
\end{equation}
Since $\Lambda_k$ is again a channel it follows that $\rho_k^{\phi}$ may depend on $\phi$ only via
linear functions of $e^{i n \phi}$, where again $n \in \{-N,\dots,N\}$. Notice that since cloning is
to preserve some information on the input state, the output state of each clone $\rho_k^{\phi}$ has
to depend on $\phi$. Since the matrix of each clone $\rho_k^{\phi}$ include at least terms $e^{\pm i
  \phi_j}$ (or possibly higher powers of these), then it follows that $\bigotimes_{k=1}^M
\rho_k^{\phi}$ contains entries that depend on $\phi$ via terms $e^{\pm i \bar{M} \phi}$ where
$\bar{M} \geq M > N$.

This leads to a contradiction, since for decorrelation to be successful
we would need the equality of a polynomial in $e^{i n \phi}$, where
$-N<n<N$, with a polynomial containing higher powers (at least $M$) of
$e^{\pm i \phi}$, and this is impossible to hold for a continuous range of
parameters $\phi$.  Hence, approximate cloning with decorrelated
clones is impossible for any set of pure states which contains a
finite arch of states of the form (\ref{phasestat}). This no
go-theorem clearly can be extended to any set of mixed states
containing an arch of the form
\begin{equation}\label{archrho}
  \rho_\phi:=U_\phi\rho U_\phi^\dag.
\end{equation}
In fact, an arch of mixed states $\rho_\phi$ can be obtained as
$\rho_\phi=\map{N}(\ket{\phi}\bra{\phi})$ with $\map{N}$ amplitude-damping channel
$\map{N}(\rho)=\alpha\rho+\beta\sigma_z\rho\sigma_z$. Therefore, if a map $\map{D}$ is able to clone
an arch of $\rho_\phi$ without correlations, then the map $\map{D}\circ\map{N}$ would do the same
for an arch of pure states, which contradicts our previous result. We have then proved that in
finite dimension any set of mixed states containing an arch of states of the form (\ref{archrho})
cannot be cloned without correlations in any approximate and asymmetric way.  This is clearly true,
as a special case, for covariant universal cloning, or any other covariant cloning of symmetric sets
of input states, for groups containing $U(1)$ as a subgroup.  Notice that in our derivation we have
used only linearity of the transformation and we have not used the trace preserving condition. This
implies that cloning without correlations is impossible also probabilistically.

The present no-cloning-without-correlation result is already quite general, however, it is likely to
be of even larger validity. We conjecture that it holds more generally for linearly dependent sets
of states. Such conjecture is supported by the fact that linearly independent states can be
probabilistically perfectly cloned \cite{duan}, so if we consider e.g.  $N$ copies of an unknown
qubit state, nothing forbids cloning without correlations for $N+1$ different qubit states, since
$\ket{\phi}\bra{\phi}^{\otimes N}$ will be linear independent states.

\section{Decorrelation for continuous variables}
We consider now the case of decorrelation for qumodes. For a couple of
qumodes in a joint seed state $\rho_{AB}$ the information
$(\alpha,\beta)$ (with $\alpha$ and $\beta$ complex) is encoded as
follows
\begin{equation}\label{qstate}
  D(\alpha )\otimes D(\beta)\rho_{AB} D(\alpha )^\dagger \otimes D(\beta)^\dagger,
\end{equation}
$D(z)=\exp(za^\dag-z^*a)$ for $z\in{\mathbb C}$ denoting a single-mode
displacement operator, $a$ and $a^\dag$ being the annihilation and
creation operators of the mode.  Here we show that it is always
possible to decorrelate any joint state of the form (\ref{qstate}),
with $\rho _{AB}$ representing a two-mode Gaussian state, namely
\begin{eqnarray}
  \rho_{AB} = \frac {1}{\pi ^2}\int d^4\vec{q} \,e^{-\frac 12 \vec{q}^T\vec{M}\vec{q}}D(\vec{q})\;,
\end{eqnarray}
where $\vec{q}=(q_1,q_2,q_3,q_4)$, $D(\vec{q})=D(q_1 +i q_2)\otimes
D(q_3 +i q_4)$, and $\vec{M}$ is the $4 \times 4$ (real, symmetric,
and positive) correlation matrix of the state, that satisfies the
Heisenberg uncertainty relation \cite{simon} $\vec{M} + \frac i 4
\vec{\Omega }\geq 0$, with $\vec {\Omega }=\oplus _{k=1}^2 \vec{\omega
}$ and $\vec{\omega }= \left (\begin{array}{cc} 0 & 1\\ -1 &0
\end{array}
\right )$.

A Gaussian decorrelation channel covariant under $D(\alpha)\otimes
D(\beta)$ is given by
\begin{equation}\label{qdecorr}
  \map{D}(\rho)= \frac {\sqrt{\hbox{det}\vec{G}}}{(2\pi)^2}\int d^4
  \vec{x}\,e^{-\frac 12\vec{x}^T\vec{G}\vec{x}} D(\vec{x}) \rho D^\dag (\vec{x}),
\end{equation}
with positive matrix $\vec G$. For suitable $\vec{G}$, the resulting
state $\map{D}(\rho _{AB})$ is still Gaussian, with a new
block-diagonal covariance matrix $\widetilde{\vec M}$, thus
corresponding to a decorrelated state.

In fact, it is easily seen that the map $\map{D}$ is covariant. Using
the relation
\begin{eqnarray}
  D(\vec{x})D(\vec{q})D(\vec{x})=
  e^{2i(q_1 x_2 -q_2 x_1+q_3 x_4 -q_4 x_3)}\, D(\vec{q})
  \;,
\end{eqnarray}
explicitly one has
\begin{eqnarray}
  &&\map{D}(\rho _{AB}) =
  \frac {\sqrt{\hbox{det} \vec{G}}}{(2\pi)^2 \pi ^2}\int d^4 q \,
  e^{-\frac 12 \vec{q}^T \vec{M}\vec{q}}\,D(\vec{q}) \nonumber \\& &
  \times \int d^4 x\,
  e^{-\frac 12 (\vec{q}\oplus \vec{x})^T \vec{G'} (\vec{q}\oplus \vec {x})}
  \;,\label{expl}
\end{eqnarray}
where $\vec{G'}$ is the $8 \times 8 $ block matrix
\begin{eqnarray}
  \vec{G'} = \left (
\begin{array}{cc}
  0  & \vec{\Sigma }^T  \\ \Vec{\Sigma }& \vec{G}
\end{array}
\right )\;,
\end{eqnarray}
with
\begin{eqnarray}
  \vec{\Sigma }= \left (
\begin{array}{cc}
  \sigma _y  & 0 \\ 0 & -\sigma _y
\end{array}
\right )\;,
\end{eqnarray}
and $\sigma _y$ denoting the usual Pauli matrix $\sigma _y= \left (
\begin{array}{cc}
0  & -i  \\ i & 0
\end{array}
\right )$.  Notice also that $\vec{\Sigma }^T =-\vec{\Sigma }$.
\par The integral on $x$ in Eq. (\ref{expl}) can be performed, and one
obtains
\begin{eqnarray}
  \map{D}(\rho _{AB})=
  \frac {1}{\pi ^2}\int d^4q \, e^{-\frac 12 \vec{q}^T (\vec{M} +
    \vec{U}) \vec {q}}\,D(\vec{q})\;,
\end{eqnarray}
where $\vec{U}=\vec{\Sigma }\vec {G}^{-1}\vec{\Sigma }$.  Then, by
writing the correlation matrix $\vec {M}$ of the input seed state in
block-form, namely
\begin{eqnarray}
  \vec{M}= \left (
\begin{array}{cc}
  \vec{A} & \vec{C} \\ \vec{C}^T & \vec{B}
\end{array}
\right )\;,
\end{eqnarray}
and writing $\vec{G}^{-1}$ as
\begin{eqnarray}
  \vec{G}^{-1}= \left (
    \begin{array}{cc}
      \vec{W} & \vec{V} \\ \vec{V}^T & \vec{Z}
\end{array}
\right )\;
\end{eqnarray}
a decorrelation map is obtained just by taking
\begin{eqnarray}
  \vec{V}= \sigma _y \vec{C} \sigma _y  \;.
\end{eqnarray}
Since for physical maps one must have $\vec{G}^{-1} >0$, then
$\vec{W}$ and $\vec{Z}$ are subject to constraints.  Typically, one
will take $\vec{W}$ and $\vec{Z}$ such that $\vec{G}^{-1} >0$ and the
added noise is minimal.  Since the channel in Eq.~\eqref{qdecorr} is
covariant also for $D(\alpha)^{\otimes 2}$, notice that the above
derivation holds for the case of encoding with the same unitary on
both qumodes as well.

In the following we will give two relevant examples of decorrelation
maps for Gaussian states.

\setcounter{subsubsection}{0}
\subsubsection{Decorrelating twin-beam states}
A special example of Gaussian state of two qumodes is the so-called
{\em twin beam}, which is an entangled state that can be generated in
a quantum optical lab by parametric down-conversion of vacuum. On the
Fock basis $\{|n \rangle \}$, this state can be written as
\begin{eqnarray}
  |\psi \rangle = \sqrt{1- \lambda ^2}\sum _{n=0}^\infty \lambda ^n
  |n\rangle \otimes |n\rangle \;,
\end{eqnarray}
with $0\leq \lambda <1$, and the correlation matrix $\vec M$ for $\rho
_{AB}= |\psi \rangle \langle \psi |$ is given by
\begin{eqnarray}
  \vec M= \frac {1+ \lambda ^2}{1- \lambda ^2} \openone - \frac {2 \lambda }{1-
    \lambda ^2} \left (
\begin{array}{cc}
  0 & \sigma _z \\ \sigma _z &0
\end{array}
\right )\;.
\end{eqnarray}
For any state in the set (\ref{qstate}), the covariant map
(\ref{qdecorr}) with
\begin{eqnarray}
  \vec G ^{-1}= \frac {2 \lambda }{1- \lambda ^2}
  \left [(1+\varepsilon )\openone + \left (
      \begin{array}{cc}
0  & \sigma _z \\ \sigma _z & 0
\end{array}
\right )\right ]\;,
\end{eqnarray}
and arbitrary $\varepsilon >0$, provides two decorrelated states,
independently of the signal $(\alpha ,\beta )$. The covariance matrix
of the decorrelated seed state is $\widetilde{\vec M} = (\frac{1+
  \lambda }{1- \lambda }+\varepsilon ') \openone $, with $\varepsilon
'= \frac{2 \lambda \varepsilon}{ 1- \lambda ^2}$, which correspond to
two thermal states with mean photon number $\bar n= \frac{\lambda }{1-
  \lambda }+\frac {\varepsilon '} 2 $ each.

\subsubsection{Decorrelating classically correlated coherent states}
Coherent states that are classically correlated via a Gaussian
function are given by the set (\ref{qstate}), where the seed state is
written as
\begin{eqnarray}
  \rho _{AB}= \int \frac{d^2 \gamma }{\pi \delta ^2} e^{-\frac {|\gamma
      |^2}{\delta ^2}} |\gamma \rangle \langle \gamma | ^{\otimes 2}
  \;,
\end{eqnarray}
and $| \gamma \rangle $ are coherent states. This seed state can be
easily obtained by mixing a thermal state with mean photon number
$\bar n= 2 \delta ^2$ with the vacuum in a $50/50$ beam splitter. The
corresponding correlation matrix $\vec{M}$ is given by
\begin{eqnarray}
  \vec{M}= (1+ 2 \delta ^2) \openone +2 \delta ^2
  \left (
\begin{array}{cc}
  0 & \sigma _z \\ \sigma _z &0
\end{array}
\right )\;.
\end{eqnarray}
A decorrelating map is obtained from (\ref{qdecorr}) with
\begin{eqnarray}
  \vec G ^{-1}= 2 \delta ^2
  \left [(1+\varepsilon )\openone - \left (
\begin{array}{cc}
  0  & \sigma _z \\ \sigma _z & 0
\end{array}
\right )\right ]\;,
\end{eqnarray}
and arbitrary $\varepsilon >0$. For any state in the set
(\ref{qstate}), such a covariant map provides two decorrelated states,
independently of the signal $(\alpha ,\beta )$. The covariance matrix
of the decorrelated seed state is $\widetilde{\vec M} = (1+ 4 \delta
^2 +\varepsilon ') \openone $, with $\varepsilon '= 2 \delta ^2
\varepsilon $, which correspond to two factorized thermal states with
mean photon number $\bar n= 2 \delta ^2 +\frac {\varepsilon '}2 $
each.

\subsection{Relation with cloning of continuous variables}
\par The striking difference between the qubit and the qumode cases is
that for qubits only few states can be decorrelated, whereas for
qumodes any joint Gaussian state can be decorrelated. This is due to
the fact that the covariance group for qubits comprises all local
unitary transformations, whereas for qumodes includes only local
displacements, which is a very small subset of all possible local
unitary transformations in infinite dimension. In particular, unlike
the case of qudits, it can be shown that states obtained via Gaussian
cloning of continuous variables can be decorrelated and the no-go
proof valid for finite dimensional cases does not apply here.
\par Cloning for continuous variables with minimal added noise can be
obtained from $N$ to $M$ copies both for coherent states \cite{clon}
and mixed states \cite{broad} as follows: 1) use a $N$-splitter which
concentrates the signal in one mode and discards the other $N-1$
modes; 2) amplify the signal by a phase-insensitive amplifier with
power gain $G=\frac MN$; 3) distribute the amplified mode by mixing it
in an $M$-splitter with $M-1$ vacuum modes.  The noise in each mode
$a_i$ is evaluated by the sum of variances $\Delta x_i ^2 +\Delta
y_i^2$ of conjugated quadratures $x_i=\frac{a_i + a_i ^\dag}{2}$ and
$y_i=\frac{a_i - a_i ^\dag}{2i}$. Notice that for Heisenberg relations
necessarily one has $\Delta x_i ^2 +\Delta y_i^2 \geq \frac 12$.  In
the concentration stage the $N$ modes with amplitude $\langle a_i
\rangle =\alpha $ and noise $\Delta x_i^2 +\Delta y_i^2 =\gamma_i$ are
reduced to a single mode with amplitude $\sqrt N \alpha $ and noise
$\gamma =\frac 1N \sum _{i=0}^{N-1}\gamma _i$.  The amplification
stage gives a mode with amplitude $\sqrt M \alpha $ and noise $\gamma
' =\frac MN \gamma +\frac {M}{2N} -\frac 12$.  Finally, the
distribution stage gives $M$ modes, with amplitude $\alpha $ and noise
$\Gamma = \frac 1M \left (\gamma ' +\frac {M-1}{2}\right )$ each. The
distribution stage produces highly correlated copies. The correlated
clones of coherent states and displaced thermal states can be simply
decorrelated as follows. First, apply the inverse transformation of
the distribution stage, retaining just the copy with amplitude $\sqrt
M \alpha$, and then 4) distribute by mixing in an $M$-splitter with
$M-1$ modes in thermal states with noise $\gamma '$ (corresponding to
mean photon number $\bar n= \gamma ' -\frac 12$). In such a way,
continuous variables clones will be decorrelated.  Clearly, the
concatenation of stages 1), 2) and 4) gives directly a $N$-to-$M$
continuous variables covariant cloning without correlation for
coherent states and displaced thermal states.

\section{Conclusions}
We addressed the problem of removing correlation from sets of states
while preserving as much local quantum information as possible.  We
reviewed the problem of decorrelation for two qubits and provided sets
of decorrelable states and the minimum amount of noise to be added for
decorrelation. In continuous variables, we showed that an arbitrary
set of bipartite Gaussian state can be decorrelated in a covariant way
with respect to the group of displacement operators, i.e.
independently of the coherent signal.  The striking difference between
the qubit and the qumode cases is that for qubits only few states can
be decorrelated, whereas for qumodes any joint Gaussian state can be
decorrelated. This is due to the fact that the covariance group for
qubits comprises all local unitary transformations, whereas for
qumodes includes only local displacements, which is a very small
subset of all possible local unitary transformations in infinite
dimension. Indeed, for the same reason decorrelation becomes much
easier when considering covariance with respect to unitary
transformations of the form $U\otimes U$ (i.~e. with the same
information encoded on the quantum systems, e.~g.  the qubit Bloch
vectors have the same direction, or the qumodes are displaced in the
same direction), which is actually the case when considering
broadcasted states. Covariant decorrelation of this kind for multiple
copies gives insight into the problem of how much individual
information can be preserved, while all correlations between copies
are removed.  As a rule of thumb, for covariant sets of states we can
say that only a small subset of states can be decorrelated if the set
is too large.

We proved that states obtained from universal cloning can only be
decorrelated at the expense of a complete erasure of local information
(i.e. information about the copied state). More generally, we proved
that cloning without correlations among the copies is impossible for
sets of qudits that contain phase-set of states. In infinite
dimension, on the contrary, we showed that it is possible to realize
continuous variable cloning without correlation between the copies, by
slightly modifying the set-up of the customary cloning of coherent
states. Among the open problems for future work, we notice that we
didn't provide any experimental scheme for covariant decorrelation,
even for two qubits. Moreover, in the case of continuous variables, we
just gave a covariant channel for decorrelation, without facing the
problem of minimizing the noise added to the output decorrelated
states. Finally, it would be interesting to prove or disprove our
conjecture about discrete set of states, namely that cloning without
correlations is impossible for linear dependent set of states.

\par The problem of removing correlations from sets of states while
preserving local information can be seen as the simplest version of a
{\em quantum cocktail-party} problem \cite{qph}.  In general, such a
problem can be formulated as follows.
Assume we have a bipartite quantum system (e.g. two qubits, two
quantum modes of electromagnetic field, etc.) initially in a state
$\ket{0}\otimes \ket{0}$ (or more generally in some mixed state
$\rho_{AB}$). The signal is encoded using unitary operations $U_A(t)$,
$U_B(t)$ acting locally at time $t$ on subsystems $A$ and $B$,
respectively. The communication of quantum signals will amount to
sending the states $[U_A(t)\otimes U_B(t)]\ket{0}\otimes \ket{0}$ at
different times $t$, each time rotated by a different pair of unitary
matrices $U_A(t)$ and $U_B(t)$, depending on the quantum message
intended to be transmitted.  After this encoding, the systems pass
through the environment which causes the two signals to be mixed in
analogy to classical mixing of signals in microphones. This mixing can
be represented by a unitary operation $V$ that entangles both systems
with the environment state $\ket{E}$ as follows
\begin{equation}
\ket{\psi(t)}_{ABE}=V(U_A(t)\otimes U_B(t)\otimes I) \ket{0} \otimes \ket{0}\otimes \ket{E}.
\end{equation}
The analog of the classical cocktail-party problem \cite{cock} would be now to
determine the ``signals'' $U_A(t)$ and $U_B(t)$---or the state
$[U_A(t)\otimes U_B(t)]\ket{0}\otimes \ket{0}$---from the output state
of $AB$ only, without even knowing the interaction with the
environment $V$: this would be a strict quantum analog of \emph{blind
  independent component separation}. In this sense we would
decorrelate the signals $U_A(t)$ and $U_B(t)$.  This quantum version
of the cocktail-party problem is much harder than its classical
counterpart, for many reasons, including the no-cloning theorem, which
forbids to determine the output state from a single copy: an
approximate solution, if possible, would need at least some additional
assumptions about the time self-correlation of each separate signal,
along with the aid of a quantum memory to store the whole
time-sequence of output states of $AB$ and a full joint measurement on
the whole sequence. We posed in this paper a simpler, but a closely related problem of
decorrelating two quantum signals, in the scenario where the signals
$U_A$ and $U_B$ are encoded on a correlated state $\rho_{AB}$ as: $U_A
\otimes U_B \rho_{AB} U^\dagger_A \otimes U^\dagger _B$, but no
additional mixing operation $V$ is applied.  We wanted to decorrelate
the received state, and the desired result is two completely
uncorrelated systems $A$ and $B$, each one in a state that carries
information about the signals $U_A$ and $U_B$, respectively.

\begin{acknowledgments}
RDD acknowledges support from the European Commission under the Integrated Project
QAP (Contract No. 015848).
\end{acknowledgments}


\begin{thebibliography}{99}
\bibitem{note:Haar} Throughout the paper we will
  consider only groups that have invariant measure (so-called ``unimodular''), and
$F(\sigma,\tau):=\Tr[(\sqrt\sigma \tau\sqrt\sigma)^\frac12]$ is the Uhlmann fidelity.
\end{thebibliography}
\end{document}